 \documentclass[preprint,5p,times,twocolumn]{elsarticle}

\usepackage{amssymb}
\usepackage{amsmath}
\usepackage[margin=0pt,font=small,labelfont=bf,labelsep=colon]{caption}

\usepackage{algorithmic}
\usepackage{algorithm}

\usepackage{graphicx}
\usepackage{amsmath}
\usepackage{color}
\usepackage{caption}
\usepackage{subcaption}

\usepackage{listings}
\usepackage{color}
\usepackage{textcomp}
\definecolor{listinggray}{gray}{0.9}
\definecolor{lbcolor}{rgb}{0.9,0.9,0.9}
\journal{Computer Physics Communications}

\begin{document}

\begin{frontmatter}

%% Title, authors and addresses

%% use the tnoteref command within \title for footnotes;
%% use the tnotetext command for the associated footnote;
%% use the fnref command within \author or \address for footnotes;
%% use the fntext command for the associated footnote;
%% use the corref command within \author for corresponding author footnotes;
%% use the cortext command for the associated footnote;
%% use the ead command for the email address,
%% and the form \ead[url] for the home page:
%%
%% \title{Title\tnoteref{label1}}
%% \tnotetext[label1]{}
\author{T. Siro\corref{cor1}}
\ead{topi.siro@aalto.fi}
\cortext[cor1]{Corresponding author}
\author{A. Harju}

%% \ead[url]{home page}
%% \fntext[label2]{}
%% \cortext[cor1]{}
%% \address{Address\fnref{label3}}
%% \fntext[label3]{}

\title{Exact diagonalization of quantum lattice models on coprocessors}

%% use optional labels to link authors explicitly to addresses:
%% \author[label1,label2]{<author name>}
%% \address[label1]{<address>}
%% \address[label2]{<address>}

\address{Aalto University School of Science, P.O. Box 14100, 00076 Aalto, Finland}

\begin{abstract}
%% Text of abstract
We implement the Lanczos algorithm on an Intel Xeon Phi coprocessor and compare its performance to a multi-core Intel Xeon CPU and an NVIDIA graphics processor. The Xeon and the Xeon Phi are parallelized with OpenMP and the graphics processor is programmed with CUDA. The performance is evaluated by measuring the execution time of a single step in the Lanczos algorithm. We study two quantum lattice models with different particle numbers, and conclude that for small systems, the multi-core CPU is the fastest platform, while for large systems, the graphics processor is the clear winner, reaching speedups of up to 7.6 compared to the CPU. The Xeon Phi outperforms the CPU with sufficiently large particle number, reaching a speedup of 2.5.
\end{abstract}

\begin{keyword}
%% keywords here, in the form: keyword \sep keyword

%% MSC codes here, in the form: \MSC code \sep code
%% or \MSC[2008] code \sep code (2000 is the default)
Tight binding \sep Hubbard model \sep exact diagonalization \sep GPU \sep CUDA \sep MIC \sep Xeon Phi
\end{keyword}

\end{frontmatter}

%%
%% Start line numbering here if you want
%%
% \linenumbers

%% main text
\section{Introduction}
\label{intro}
In recent years, there has been tremendous interest in utilizing coprocessors in scientific computing, including condensed matter physics\cite{Intro1,Intro2,Intro3,GPUreview}. Most of the work has been done on graphics processing units (GPU), resulting in impressive speedups compared to CPUs in problems that exhibit high data-parallelism and benefit from the high throughput of the GPU. In 2013, a new type of coprocessor emerged on the market, namely the Xeon Phi by chip manufacturer Intel. The Xeon Phi is based on Intel's many integrated core (MIC) architecture, and features around 60 CPU cores that can be easily programmed with existing paradigms, such as OpenMP and MPI. The performance of the Xeon Phi has also already been investigated in some computational physics research areas with mixed results in comparison to GPUs.\cite{IntroMIC1,IntroMIC2,IntroMIC3}.

In this work, we apply the Xeon Phi coprocessor to solving the ground state energy of a quantum lattice model by the Lanczos algorithm and compare its performance to a multi-core CPU and a GPU. Previously, the Lanczos algorithm has been implemented on a GPU with speedups of up to around 60 and 100 in single and double precision arithmetic, respectively, in comparison to a single-core CPU program\cite{Siro_2012}.

We examine the tight binding Hamiltonian
\begin{eqnarray}
\displaystyle H &=& -t\sum_{<ij>}\sum_{\sigma=\uparrow,\downarrow}(c_{i,\sigma}^{\dagger}c_{j,\sigma}+h.c),
\label{eq:H}
\end{eqnarray}
where $<ij>$ denotes a sum over neighboring lattice sites, $c_{i,\sigma}^{\dagger}$ and $c_{i,\sigma}$ are the creation and annihilation operators which respectively create and annihilate an electron at site $i$ with spin $\sigma$, and $n_{i,\sigma}=c_{i,\sigma}^{\dagger}c_{i,\sigma}$ counts the number of such electrons. The hopping amplitude is denoted by $t$. The tight-binding model describes free electrons hopping around a lattice, and it gives a crude approximation of the electronic properties of a solid. The model can be made more realistic by adding interactions, such as on-site repulsion, which results in the well-known Hubbard model\cite{Hubbard}. In our basis, however, such interaction terms are diagonal, rendering their effect on the computational complexity insignificant when we consider operating with the Hamiltonian on a vector. The results presented in this paper therefore apply to a wide range of different models. 

We will solve the lowest eigenvalue, i.e. the ground state energy, of the Hamiltonian numerically with the exact diagonalization (ED) method. This simply means forming the matrix representation of $H$ in a suitable basis and using the Lanczos algorithm to accurately compute the ground state energy. The major advantage of this method is the accuracy of the results, which are essentially exact up to the numerical accuracy of the floating point numbers. The downside is that using the full basis is very costly, since its size scales exponentially with increasing system size and particle number. This means that we are limited to quite small systems. Despite this limitation, the ED method has been successful in many very topical areas of physics, including e.g. the topological properties of condensed matter systems\cite{Sheng_2011,Yang_2012,Siro_2014}.

\section{Exact diagonalization}
\subsection{The Hamiltonian}
\label{Hamiltonian}

In a lattice with $N_s$ sites with $N_{\uparrow}$ spin up electrons and $N_{\downarrow}$ spin down electrons, the dimension of the Hamiltonian is just the number of ways of distributing the electrons into the lattice, taking into account the Pauli exclusion principle that forbids two or more electrons of the same spin from occupying the same site. Thus, the dimension is
\begin{equation}
\dim H = \binom{N_{s}}{N_{\uparrow}}\binom{N_{s}}{N_{\downarrow}}.
\end{equation}
The size of the basis grows extremely fast. For example, in the half-filled case where $N_{\uparrow} = N_{\downarrow} = N_{s}/2$, for 12 sites $\dim H=853776$, for 14 sites $\dim H\approx11.8\times10^{6}$ and for 16 sites $\dim H\approx166\times10^{6}$. In addition, the matrices are very sparse, because the number of available hops, and thus the number of nonzero elements in a row, grows only linearly while the size of the matrix grows exponentially.

\begin{figure}[t]
\includegraphics[width=0.35\textwidth]{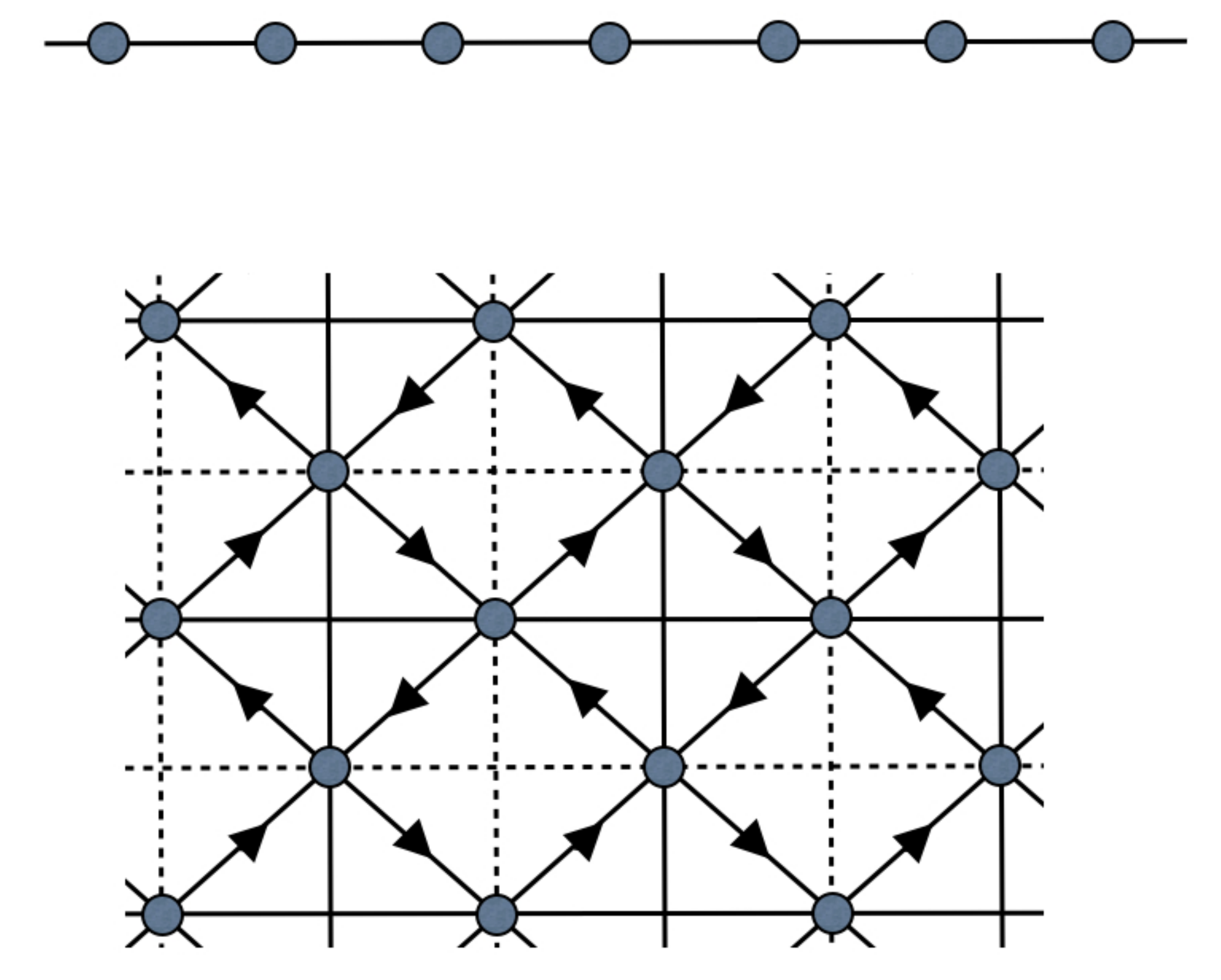}
\caption{\label{figlattices} The two lattice geometries. (top) A 1D lattice with nearest neighbor hoppings. (bottom) A checkerboard lattice with complex nearest-neighbor hoppings (the arrows indicate the sign of the complex phase), real next-nearest neighbor hoppings with alternating sign (indicated by the dashed and solid lines) and real third nearest-neighbour hoppings (not drawn for clarity).}
\end{figure}

We study two different lattice geometries, presented in Figure \ref{figlattices}. The first is a simple 1-dimensional lattice with nearest-neighbour hoppings. We use a lattice with 26 and 18 sites for the one and two spin species cases, respectively. The other is a checkerboard lattice introduced in Reference \cite{Sun_2011}. It is a widely studied lattice, because with a nearest-neighbor interaction, an analogue to the fractional quantum Hall effect can be observed in the lattice without an external magnetic field\cite{Sheng_2011}. It also contrasts the 1D lattice because it is two-dimensional and has twelve hoppings  per site, compared to only two in the 1D lattice. This leads to a much denser hopping Hamiltonian. We use a checkerboard lattice with 30 and 18 sites for the one and two spin species cases, respectively. In all lattices, periodic boundary conditions are always used.

For a detailed description of forming and storing the Hamiltonian, see Reference \cite{Siro_2012}. A similar scheme has also been used in Reference \cite{Sharma}. To summarize, the Hamiltonian can be split into spin up and spin down parts as
\begin{equation}
H=H_{\uparrow}\otimes I_{\downarrow}+I_{\uparrow}\otimes H_{\downarrow},\label{Hsplit}
\end{equation}
where $I_{\sigma}$ is the identity operator for electrons with spin  $\sigma$
and $\otimes$ is the tensor product. The basis states for a single spin species, up or down, are represented by integers whose set and unset bits correspond to occupied and unoccupied sites, respectively. Then, the hopping Hamiltonians $H_{\uparrow}$ and $H_{\downarrow}$ are computed in the basis and stored in the memory in the ELL sparse matrix format. 

The ELL format stores a sparse matrix into two dense matrices that contain the nonzero matrix elements and the corresponding column indices. The width of the matrices is the maximum number of nonzero elements per row in the original matrix. For an example of the ELL sparse matrix format, see Figure \ref{ELL}. The nonzero density for the matrices we have used ranges from $10^{-3}$ to $10^{-6}$. We use ELL instead of other standard formats, such as CSR, because in $H$, there is quite little variation in the number of nonzeros per row. This means that we do not have to add a lot of padding zeros into the ELL format matrices. Also, in our tests, we found the performance with CSR to be essentially identical to ELL, so we use the simpler method.  

\begin{figure}[t]
\[
A=\left(\begin{array}{cccc}
5 & 1 & 0 & 0\\
0 & 2 & 7 & 3\\
4 & 0 & 6 & 0\\
0 & 9 & 8 & 0\end{array}\right)\]
\[
\Downarrow\]
\[
\text{data}=\left(\begin{array}{ccc}
5 & 1 & *\\
2 & 7 & 3\\
4 & 6 & *\\
9 & 8 & *\end{array}\right)\quad\text{indices}=\left(\begin{array}{ccc}
0 & 1 & *\\
1 & 2 & 3\\
0 & 2 & *\\
1 & 2 & *\end{array}\right)\]
\[
\Downarrow\]
\begin{eqnarray*}
\text{data} & = & (5,2,4,9,1,7,6,8,*,3,*,*)\\
\text{indices} & = & (0,1,0,1,1,2,2,2,*,3,*,*)\end{eqnarray*}
\caption{An example of using the ELL format. It produces two smaller matrices from the initial
matrix. In practice, these will be converted to vectors in column-major
order for the GPU and row-major order for the CPU and the Xeon Phi. The stars denote padding and they are set to zero.}
\label{ELL}
\end{figure}

\subsection{The Lanczos algorithm}
\label{lanczos}

Because of the very fast growth of the Hilbert space dimension as a function of the particle number, fully diagonalizing the Hamiltonian is only possible for rather small systems and with only a few particles. Usually, we are mostly interested in the smallest eigenvalues and states. These can be accurately approximated with iterative algorithms, one of which is the Lanczos algorithm\cite{itermethods}.

In the Lanczos algorithm, the Hamiltonian is projected onto an orthogonalized basis in a Krylov subspace, defined by
\begin{equation}
\mathcal{K}_{m}(f,H)=\text{span}(f,Hf,H^{2}f,\dots,H^{m-1}f),
\end{equation}
where $f$ is a random starting vector and $m$ is the Krylov space dimension. The result of the Lanczos iteration is a tridiagonal matrix, i.e. one with nonzero elements only on the main diagonal and the first sub- and superdiagonals. The dimension of the resulting matrix is equal to $m$. As $m$ increases, the lowest (and highest) eigenvalue of the matrix gives an increasingly accurate approximation of the corresponding eigenvalue of $H$. Importantly, sufficient convergence occurs typically already for $m\approx100$, even when the Hamiltonian matrix is very large.

\begin{algorithm}[t]                      % enter the algorithm environment
\caption{The Lanczos algorithm \cite{itermethods}.}          % give the algorithm a caption
\label{alg1}                           % and a label for \ref{} commands later in the document
\begin{algorithmic}[1]  
\REQUIRE a random initial vector $f_1$ of norm 1
 \STATE $b_1 \leftarrow 0$
 \STATE $f_0 \leftarrow 0$
\FOR{$j=1$ to $m$}
     \STATE $q_j \leftarrow Hf_{j}-b_{j}f_{j-1}$
     \STATE $a_j \leftarrow q_{j}^{\dagger}f_j$
     \STATE $q_j \leftarrow q_{j}-a_{j}f_j$
     \STATE $b_{j+1} \leftarrow \sqrt{q_{j}^{\dagger}q_j}$. 
     \IF{$b_{j+1}=0$} 
     \STATE Stop
     \ENDIF
     \STATE $f_{j+1} \leftarrow q_{j}/b_{j+1}$
\ENDFOR
\end{algorithmic}
\end{algorithm}

In Algorithm 1, we give the pseudocode for the Lanczos algorithm. It generates the so called Lanczos basis, $\{f_{1},f_{2},\dots,f_{m}\}$,
in the Krylov space by orthonormalizing the Krylov space basis vectors. Then, the projection of $H$ in this basis is given
by the generated constants $a_{j}$ and $b_{j}$ as\[
\label{Tmatrix}
T=\left(\begin{array}{ccccc}
a_{1} & b_{2} & 0 & \cdots & 0\\
b_{2} & a_{2} & b_{3} & \ddots & \vdots\\
0 & \ddots & \ddots & \ddots & 0\\
\vdots & \ddots & b_{m-1} & a_{m-1} & b_{m}\\
0 & \cdots & 0 & b_{m} & a_{m}\end{array}\right).\]
The eigenvalues of $T$ can then be computed easily by standard methods.

\section{Hardware environment}

In this work, we compare the performance of three systems: an Intel Xeon E5-2620v2 CPU, an NVIDIA Tesla K40 GPU and an Intel Xeon Phi 7120X coprocessor. 

GPUs have been increasingly popular in scientific computing in recent years, offering impressive speedups in data-parallel problems that can support parallelism up to tens of thousands of concurrent threads. Originally designed to output graphics, the modern GPUs can be programmed to perform general purpose computation. Most of the work has been done with the CUDA programming model and language by Nvidia. CUDA is a simple extension of the C++ programming language that allows the programmer to write special functions that execute on the GPU.

The GPU consists of multiple streaming multiprocessors (SM), each containing hundreds of cores. Every SM has an L1 cache and there is also a larger L2 cache shared by all SMs. Finally, there is the main memory, called global memory, that can be accessed from the host system via the PCIe bus. The theoretical peak performances of the Tesla K40 GPU are 5 and1.66  TFLOPS in single and double precision, respectively. It has 12 GB of memory with a 288 GB/s bandwidth. 

The Xeon Phi 7120X has 61 cores based on the x86 architecture connected by a bidirectional ring interconnect. Each core supports up to four simultaneous threads and 512-bit wide SIMD vectors, meaning that they can process sixteen single precision or eight double precision floating point numbers simultaneously. The theoretical peak performances are 2.4 and 1.2 TFLOPS in single and double precision, respectively. It has 16 GB of memory with a 352 GB/s bandwidth. This is only the first generation of Xeon Phi products and the second generation, codenamed Knights Landing, is scheduled to be released before the end of 2015.

Both the GPU and the Xeon Phi serve the same purpose, namely to speed up portions of the program that benefit from the large scale parallelization. Both are connected to the host system via the PCIe bus, so the speedup should be significant enough to overcome the performance hit from the data transfers to and from the accelerator. The Xeon Phi supports two different operating modes: offload and native. In the offload mode, the main program runs on the CPU and offloads the parallel parts onto the Xeon Phi. In the native mode, the whole program is executed on the coprocessor that is running a Linux operating system. In our case, all the steps in the Lanczos algorithm can be effectively parallelized, so we use the native mode.

The major difference between the three platforms is the degree of parallelism: while the 6-core CPU with hyper threading can run 12 concurrent threads, the Xeon Phi and the GPU support up to 244 and 2880 threads, respectively. From a programming point of view, the Xeon Phi can be thought of as a big multi-core CPU, supporting standard parallel programming paradigms such as OpenMP and MPI. This allows, at least in principle, the programmer to run existing parallel codes on the coprocessor with minimal changes to the code, or parallelize serial code with ease. On the other hand, programming GPUs requires more effort, since efficient low level programming with CUDA requires knowing the hardware with its different memories and learning the GPU specific programming techniques.

\section{Programming}
\label{pro}

We program our GPU with CUDA, a parallel computing programming model developed by NVIDIA for its GPUs. With CUDA, essentially an extension of the C language, the programmer can write special functions called kernels that run on the GPU. The kernels are executed on the GPU by threads that are organized in independent blocks. The launch configuration, i.e. the numbers of blocks and threads per block are defined when calling the kernel. The parallel code is written from the point of view of a single thread, and intrinsic variables, such as the id number of the thread within the block, are used to guide different threads to operate on different data. 

To obtain the best performance, the kernels should be programmed to utilize the so called shared memory, which is a fast memory space that can be used to communicate between threads belonging to the same block. Optimally, the threads should load the data from the global memory to the shared memory, perform the calculation and then write the result back to the global memory. One should also try to e.g. optimize the memory access patterns, i.e. to access contiguous data in the memory with contiguous threads. For a comprehensive overview of CUDA and optimization techniques, we refer to Reference \cite{CUDAguide}.

Our CPU and Xeon Phi programs are written in C++ and the parallel portions of the code utilize the OpenMP API. One of the selling points of the Xeon Phi coprocessor is the portability of existing multi-core CPU codes. In principle, a CPU code parallelized with OpenMP can be compiled to run on the coprocessor with no changes in the source code. In practice, the programmer should pay special attention to details like proper vectorization of inner loops and alignment of the memory allocations. For the benchmarks presented in this paper, the CPU and the Xeon Phi are running the same code. The code has been optimized for the Xeon Phi, but according to our experimentation, the CPU performance was largely unaffected by the optimizations. All inner loops were confirmed to be vectorized by the icpc compiler. For specifics on the optimization of Xeon Phi programs, we refer to Reference \cite{MICbook}.

\begin{algorithm}[t]                     % enter the algorithm environment
\caption{The GPU kernel pseudocode for operating with the Hamiltonian}          % give the algorithm a caption
\label{gpuspmv}                           % and a label for \ref{} commands later in the document
\begin{algorithmic}[1] 
\REQUIRE vector y initialized to 0
\REQUIRE blockID \COMMENT{the thread block index}
\REQUIRE sv \COMMENT{the subvector index}
\REQUIRE id \COMMENT{the thread index within the subvector}
\REQUIRE gid \COMMENT{the global thread id within the whole vector}
\REQUIRE blockID $<$ dimUp * blocksPerSubvector

\STATE sum $\leftarrow 0$
\STATE
\IF{threadIdx.x $<$ numcolsUp}         
          \STATE Ax$_s$[threadIdx.x]$ \leftarrow$ AxUp[threadIdx.x*dimUp + sv]
          \STATE Aj$_s$[threadIdx.x]$ \leftarrow$ AjUp[threadIdx.x*dimUp + sv]
\ENDIF     

\STATE syncthreads
\IF{ id $<$ dimDn}  	  
	  
\STATE
\FOR{$i=0$ to numcolsUp}         
      
          \STATE sum $\leftarrow$ sum + Ax$_s$[i] * x[Aj$_s$[i] * dimDn + id]
      
\ENDFOR

\FOR{$i=0$ to numcolsDn}   
      \STATE Aij $\leftarrow$ AxDn[i * dimDn + id]
      \STATE col $\leftarrow$ AjDn[i * dimDn + id]    

   %   \IF{Aij $\neq$ 0}	
         \STATE sum $\leftarrow$ sum + Aij * x[sv * dimDn + col]		
      %\ENDIF
      %\STATE syncthreads
\ENDFOR
\STATE y[gid] $\leftarrow$ sum;

\ENDIF
\end{algorithmic}
\end{algorithm}

All benchmarks are run on a single Xeon Phi and a single GPU. A multi-GPU/Phi implementation to allow studying larger systems is not feasible, since due to the exponential growth of the basis size, significantly increasing the system size is impossible due to memory constraints. For example, the current implementation can handle a system of 16 lattice sites with 8 up and 8 down spin particles. In this case, the size of a single state vector in double precision is 2.7 GB. The next largest half-filled case would be 18 sites with 9 up and 9 down spin particles. Here, the state vector already requires 38 GB of memory, which already exceeds the memory available on any coprocessor.

Furthermore, a multi GPU/Phi implementation would face significant challenges in overcoming the latency associated with communication between coprocessors. With current technology, both the GPU and the Xeon Phi can only communicate with another coprocessor through the host system via a slow PCIe bus. This would probably negate any potential benefits of multiple accelerators. However, this problem could be somewhat alleviated by new technologies, such as the NVLink interconnect, introduced for Pascal generation GPUs, which enables up to 160 GB/s bidirectional bandwidth between two GPUs\cite{Pascal}.

\begin{algorithm}[t]                     % enter the algorithm environment
\caption{The CPU and Xeon Phi pseudocode for operating with the Hamiltonian}          % give the algorithm a caption
\label{ompspmv}                           % and a label for \ref{} commands later in the document
\begin{algorithmic}[1] 
\REQUIRE vector y initialized to 0
\REQUIRE gid \COMMENT{the global thread id within the whole vector}
\STATE $\#$pragma omp parallel for
\FOR{sv=0 to dimUp}
	\FOR{i=0 to numcolsUp}
		\STATE idx $\leftarrow$ sv*numcolsUp + i
		\FOR{id=0 to dimDn}
			\STATE y[gid] $\leftarrow$ y[gid] + AxUp[idx] * x[AjUp[idx] * dimDn + id] 
		\ENDFOR
	\ENDFOR
\ENDFOR
\STATE
\STATE $\#$pragma omp parallel for
\FOR{sv=0 to dimUp}
	\FOR{row=0 to dimDn in steps of blocky}
		\FOR{col=0 to numcolsDn in steps of blockx}
			\FOR{r=row to row+blocky}
				\IF{r$<$dimDn}
					\FOR{c=col to col+blockx}
						\STATE idx $\leftarrow$ r*numcolsDn + c
						\STATE y[gid] $\leftarrow$ y[gid] + AxDn[idx] * x[AjDn[idx] + sv * dimDn]
					\ENDFOR
				\ENDIF
			\ENDFOR
		\ENDFOR
	\ENDFOR
\ENDFOR

\end{algorithmic}
\end{algorithm}

\begin{figure*}[t.]

\centering
\begin{subfigure}{0.5\textwidth}
\centering
 \includegraphics[width=0.9\textwidth]{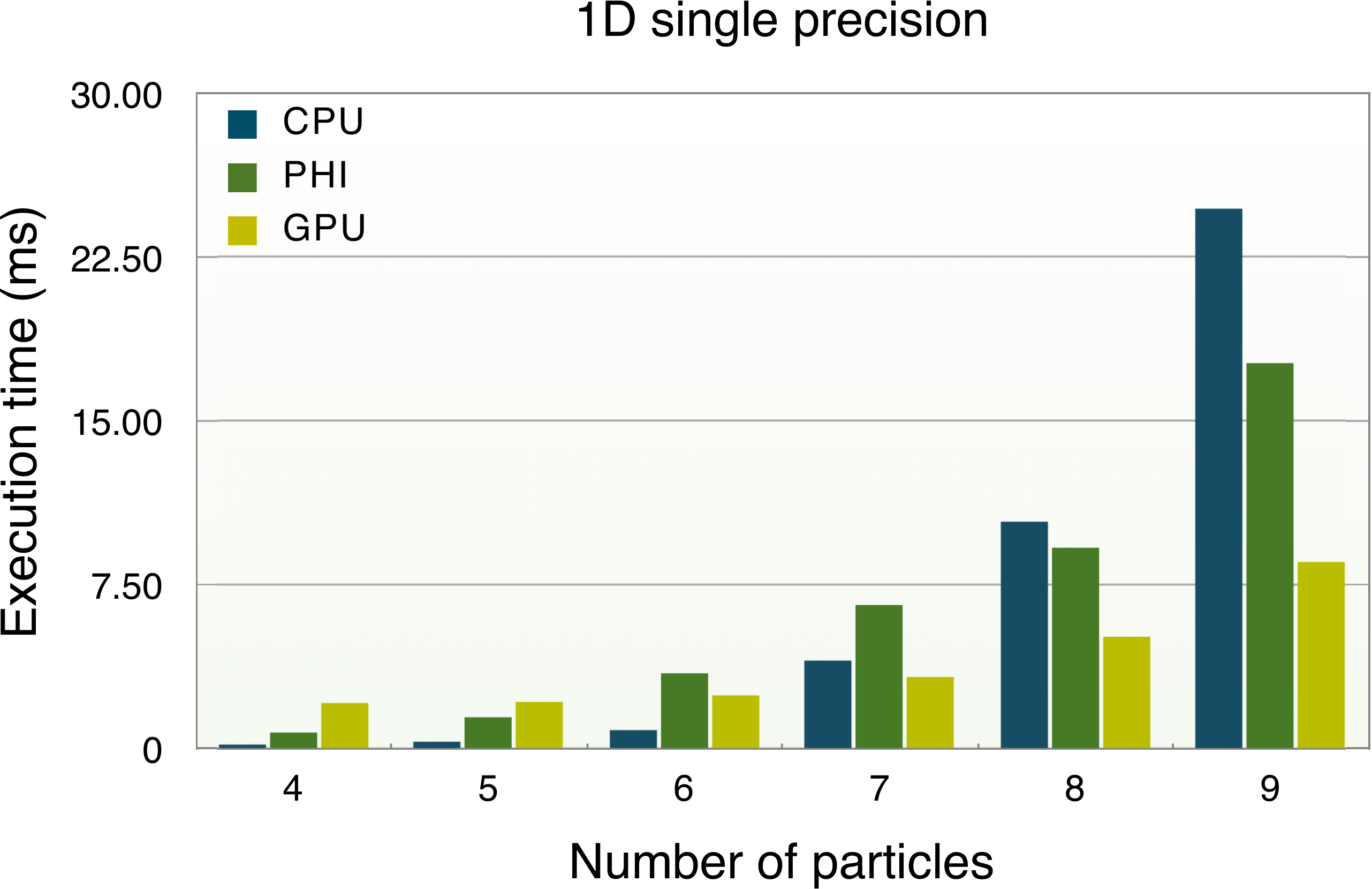}
 
 \end{subfigure}%
 \begin{subfigure}{0.5\textwidth}
 \centering
 \includegraphics[width=0.9\textwidth]{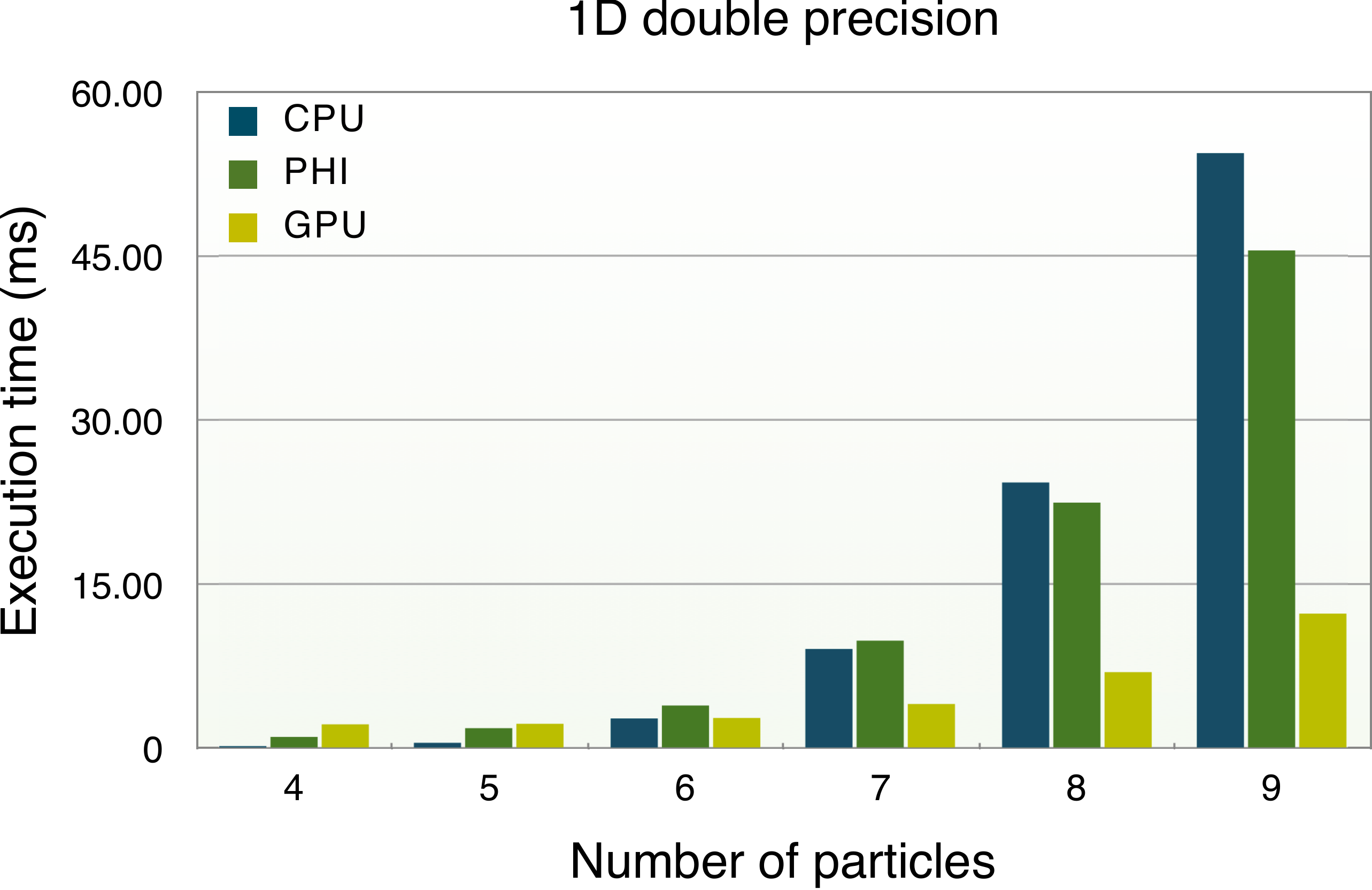}
  \end{subfigure}
  \\
  \vspace{1cm}
\begin{subfigure}{0.5\textwidth}
\centering
 \includegraphics[width=0.9\textwidth]{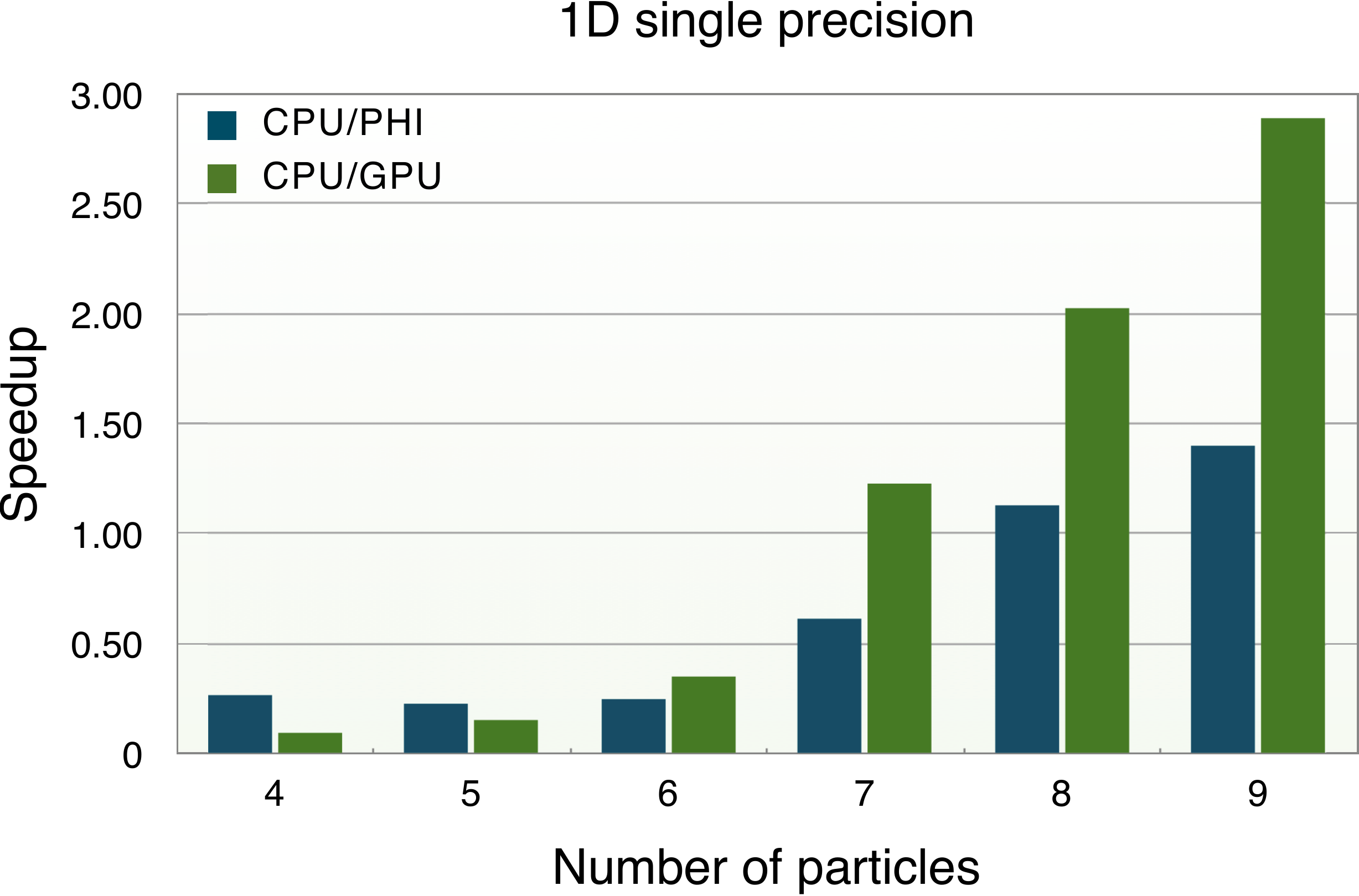}
 \end{subfigure}%
 \begin{subfigure}{0.5\textwidth}
 \hspace{0.5cm}
 \includegraphics[width=0.9\textwidth]{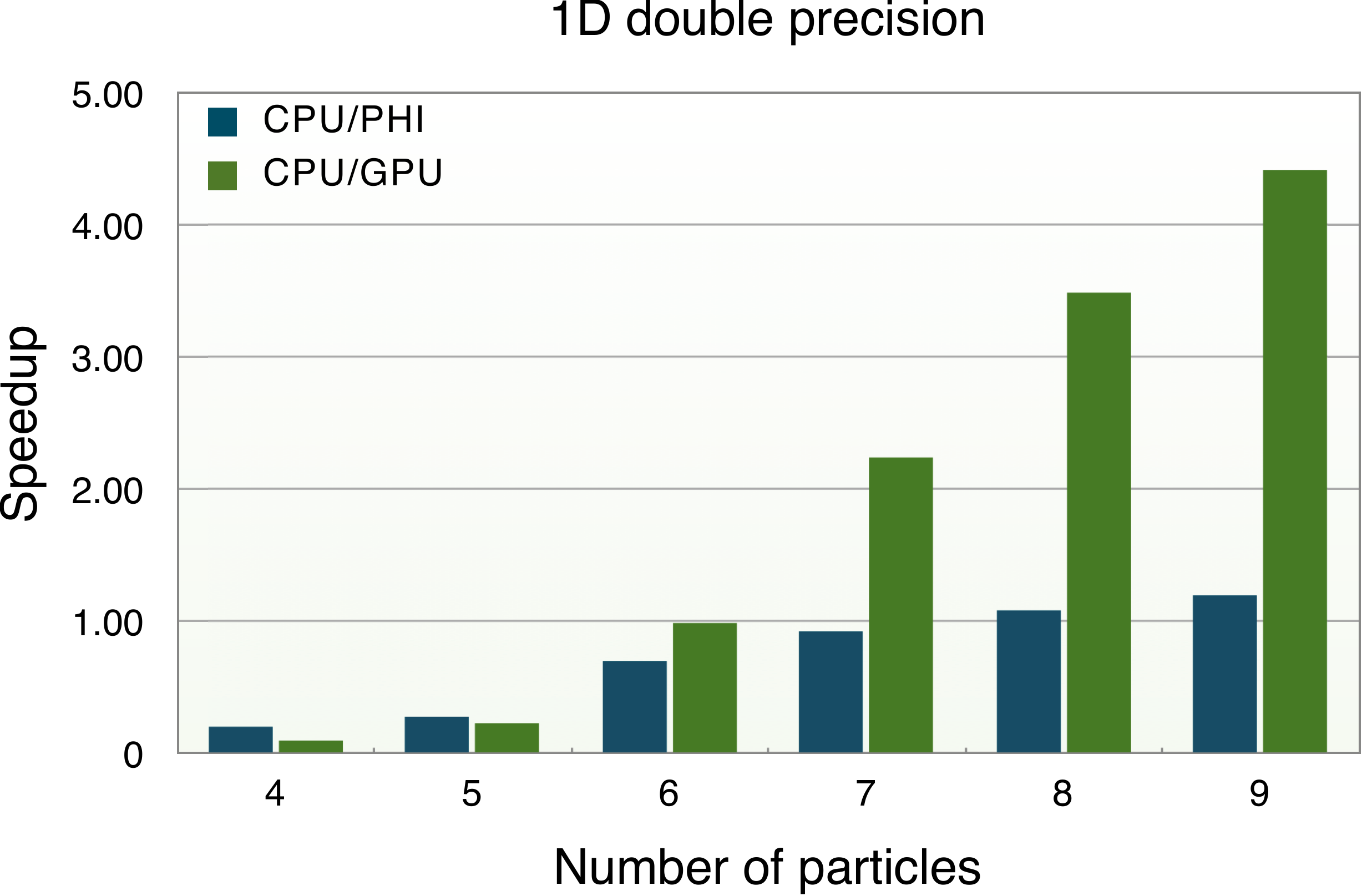}
  \end{subfigure}
  \caption{ (Color online) \label{fig1D1} (top row) The execution times of the CPU, the Xeon Phi and the GPU in the 1D lattice with different particle numbers. All particles have the same spin. (bottom row) The speedup factors of the Xeon Phi and the GPU compared to the CPU, computed from the execution times in the top row figures.}  
 \end{figure*}

 \begin{figure*}[ht.]
 
\centering
\begin{subfigure}{0.5\textwidth}
\centering
 \includegraphics[width=0.9\textwidth]{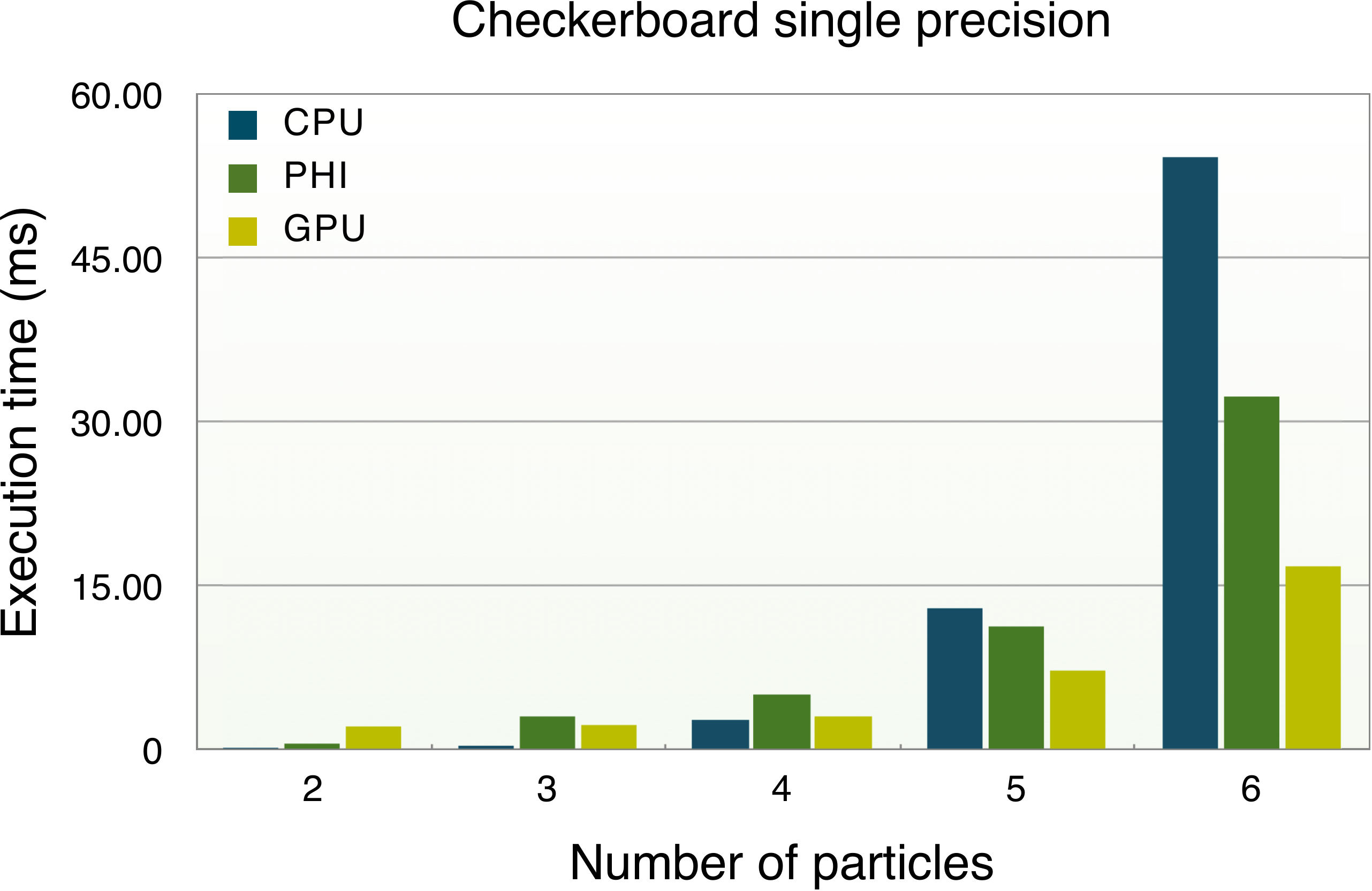}
 \end{subfigure}%
 \begin{subfigure}{0.5\textwidth}
 \centering
 \includegraphics[width=0.9\textwidth]{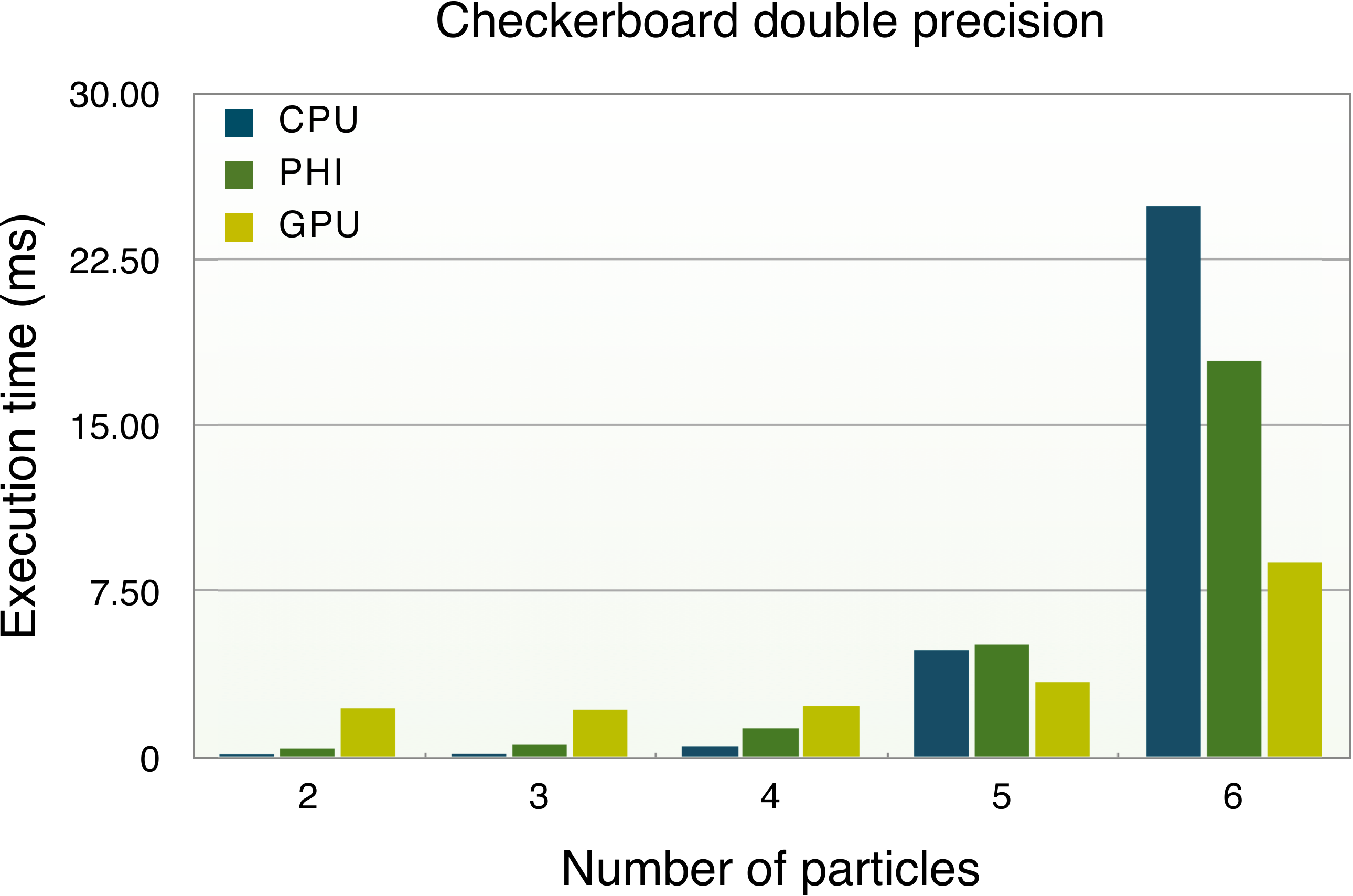}
  \end{subfigure}
  \\
  \vspace{1cm}
\begin{subfigure}{0.5\textwidth}
\centering
 \includegraphics[width=0.9\textwidth]{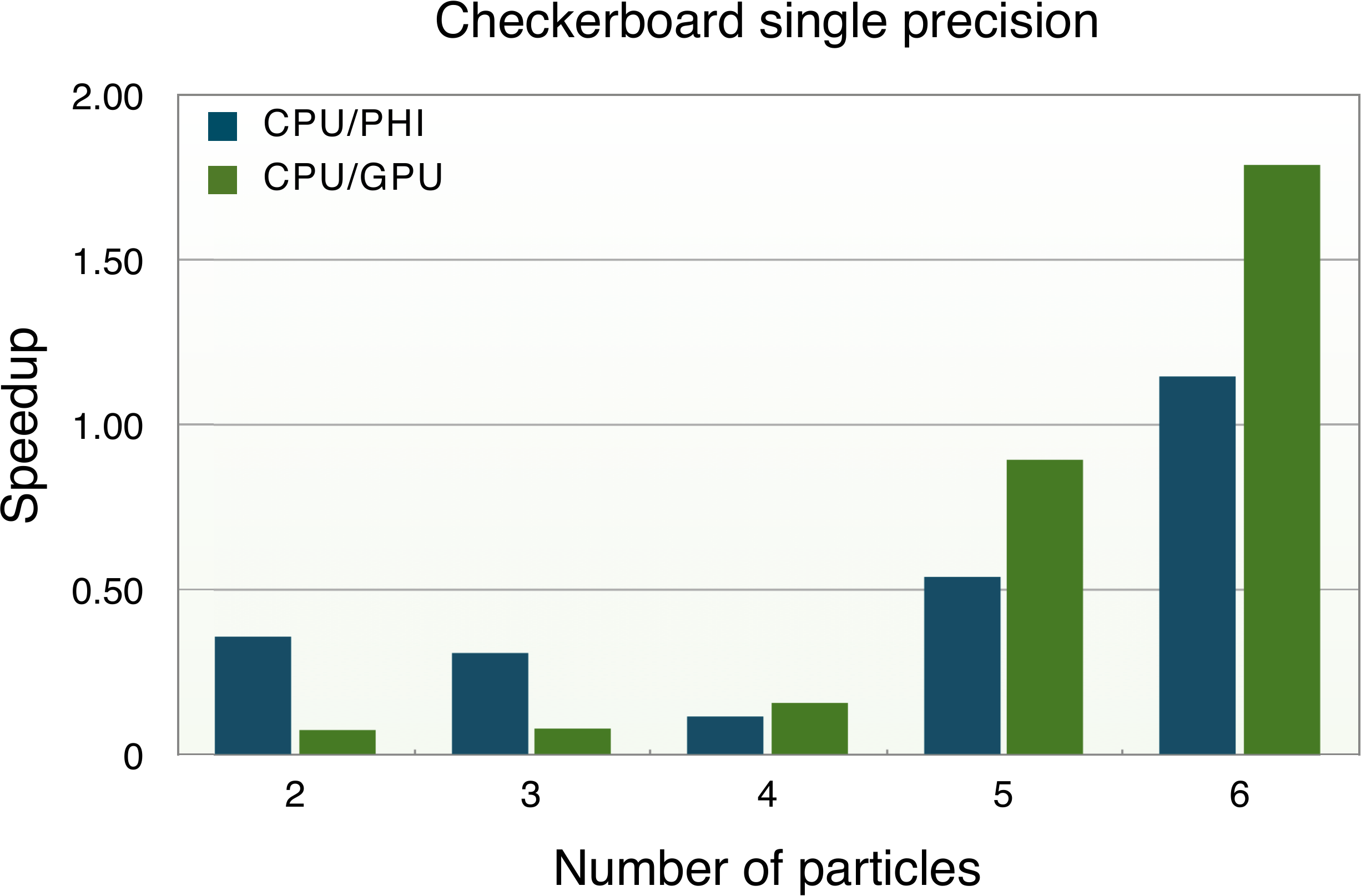}
 \end{subfigure}%
 \begin{subfigure}{0.5\textwidth}
 \hspace{0.5cm}
 \includegraphics[width=0.9\textwidth]{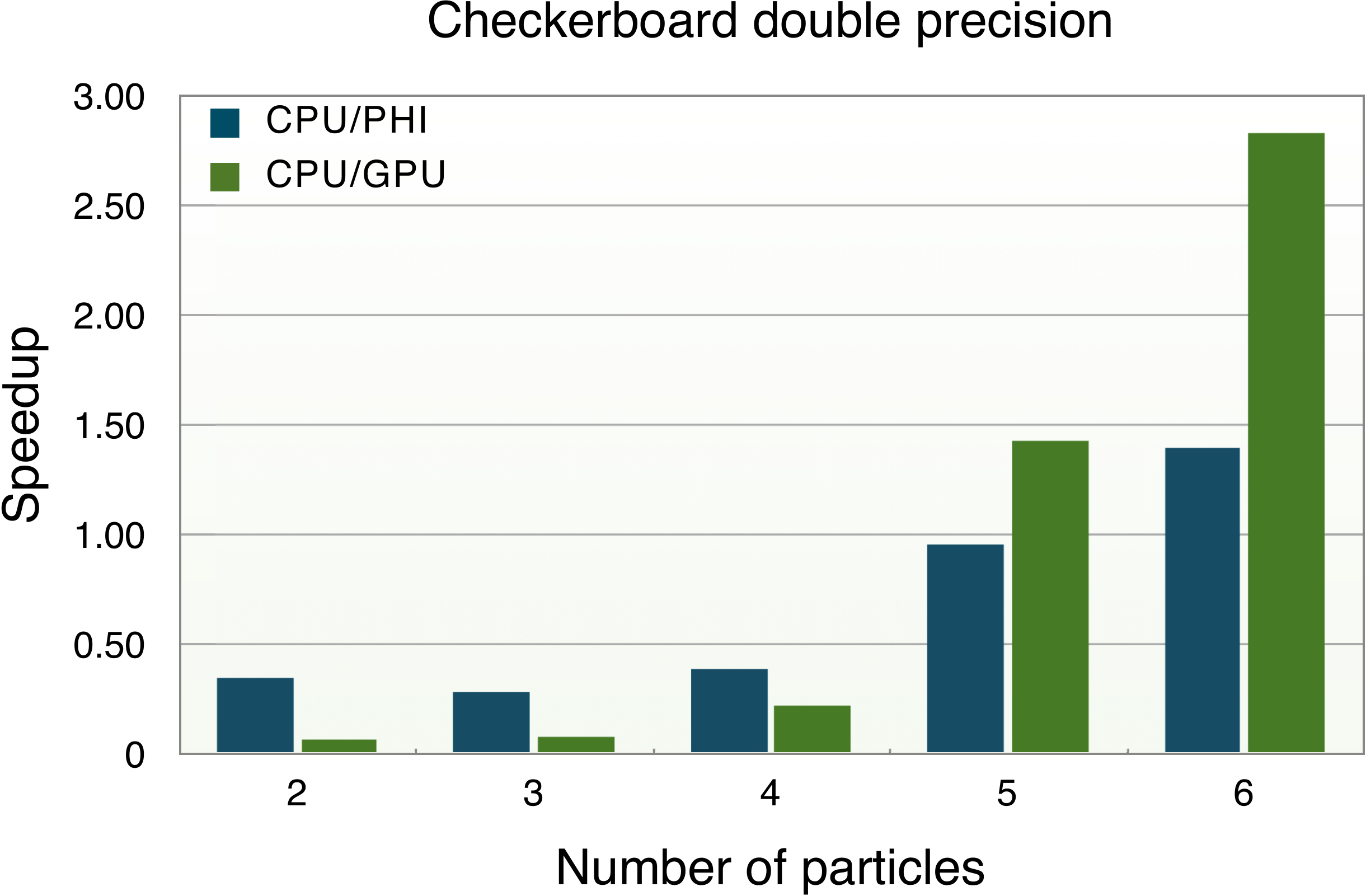}
  \end{subfigure}
  \caption{ (Color online) \label{figCh1} (top row) The execution times of the CPU, the Xeon Phi and the GPU in the checkerboard lattice with different particle numbers. All particles have the same spin. (bottom row) The speedup factors of the Xeon Phi and the GPU compared to the CPU, computed from the execution times in the top row figures.}  
 \end{figure*}

The most complicated part of the Lanczos algorithm is the sparse matrix-vector multiplication (SpMV) on line 4 of Algorithm \ref{alg1}. For large particle numbers, it is by far the most time consuming operation in the algorithm.  SpMV is a very important operation in countless areas of the computational sciences, and has thus been extensively studied. When we only have a single spin species in our system, we can form the full Hamiltonian and use an optimized library implementation for the SpMV operation. We will use the MKL library for the CPU and the Xeon Phi, and the CUSPARSE library for the GPU.

However, with two spin species, forming the full Hamiltonian matrix is out of the question for all but the very smallest of systems. Thus, we will only store the individual hopping matrices for up and down-spin electrons separately. Our GPU implementation of the Lanczos algorithm has been previously discussed in detail in Ref. \cite{Siro_2012}. For the sake of completeness, we give the pseudocode for the SpMV kernel in Algorithm \ref{gpuspmv}. In the pseudocode, Ax and Aj (with either Up or Dn as a suffix to indicate the spin) refer to the data and indices matrices in the ELL format (Figure \ref{ELL}), respectively. Further, dim and numcols with their suffixes refer to the dimension and the number of columns in the ELL matrices, respectively. The subscript $s$ refers to shared memory.

On the Xeon Phi, there is a very advanced SpMV implementation reported in Ref. \cite{MICspmv}. However, with two spin species, we do not have access to the full Hamiltonian matrix, so these techniques do not directly apply to our problem. Furthermore, keeping in mind that our GPU implementation is quite simple, we would like to keep the required programming effort comparable across the three test platforms and focus on comparing the relative performance of the systems instead of striving for the best absolute performance with complicated matrix reordering schemes.

Looking at Equation \ref{Hsplit}, the effect of operating on a state vector with $H$ can be understood by considering the vector to consist of $\dim H_{\uparrow}$ subvectors of length $\dim H_{\downarrow}$. The spin-up configuration stays constant within a subvector. The spin-up part of the Hamiltonian can then be thought to operate on a vector that consists of the subvectors: 
\begin{equation}
 (H_{\uparrow}\otimes I_{\downarrow})x=H_{\uparrow}\left(\begin{array}{c}
\mathbf{x}^{(0)}\\
\mathbf{x}^{(1)}\\
\vdots\\
\mathbf{x}^{(\dim H_{\uparrow}-1)}\end{array}\right).
\end{equation}

Correspondingly, the spin-down part of the Hamiltonian operates like a normal matrix-vector product for each of the $\dim H_{\uparrow}$ subvectors:
\begin{equation}
(I_{\uparrow}\otimes H_{\downarrow})x=\left(\begin{array}{c}
H_{\downarrow}\mathbf{x}^{(0)}\\
H_{\downarrow}\mathbf{x}^{(1)}\\
\vdots\\
H_{\downarrow}\mathbf{x}^{(\dim H_{\uparrow}-1)}\end{array}\right).
\end{equation}

Thus, for operating on a vector with the Hamiltonian, we use a straightforward implementation, where the subvectors are divided among the OpenMP threads, see Algorithm \ref{ompspmv}. For the matrix-vector products in the spin-down part, we partition $H_{\downarrow}$ into rectangular blocks of size blockx*blocky. Then, a single OpenMP thread is assigned a row of blocks and it computes the result block by block. This improves the cache usage compared to just computing the dot products of the matrix rows and the vector row by row. Experimentation showed that blockx $=16$ and blocky $=8$ gave the best performance in most cases so those are the values used in all results. 

 \begin{figure*}[t]

%\begin{subfigure}{\textwidth}
\begin{minipage}{\textwidth}
  \begin{minipage}[b]{0.5\textwidth}
    \centering
    
   \begin{tabular}{|c|c|c|c|}
   \multicolumn{4}{c}{1D single-precision execution times}\\[0.25cm]
   \hline 
\# of particles & CPU(ms) & PHI(ms) & GPU(ms)\tabularnewline
\hline
\hline 
4 & 0.31 & 1.09 & 2.18 \tabularnewline
\hline 
6 & 3.88 & 6.05 & 2.70 \tabularnewline
\hline 
8 & 70.6 & 29.2 & 10.4 \tabularnewline
\hline 
10 & 579 & 245 & 88.4 \tabularnewline
\hline 
\end{tabular}
  \end{minipage}
  \hfill
  \begin{minipage}[b]{0.5\textwidth}
    \centering
    \begin{tabular}{|c|c|c|c|}
     \multicolumn{4}{c}{1D double-precision execution times}\\[0.25cm]
\hline 
\# of particles & CPU(ms) & PHI(ms) & GPU(ms)\tabularnewline
\hline
\hline 
4 & 0.35 & 1.34 & 2.26 \tabularnewline
\hline 
6 & 7.11 & 9.00 & 3.16 \tabularnewline
\hline 
8 & 135 & 105 & 17.7 \tabularnewline
\hline 
10 & 1103 & 926 & 178 \tabularnewline
\hline 
\end{tabular}

        \end{minipage}
  \end{minipage}

 \hfill
 
  \vspace{0.6cm}
\begin{subfigure}{0.5\textwidth}
\centering
 \includegraphics[width=0.9\textwidth]{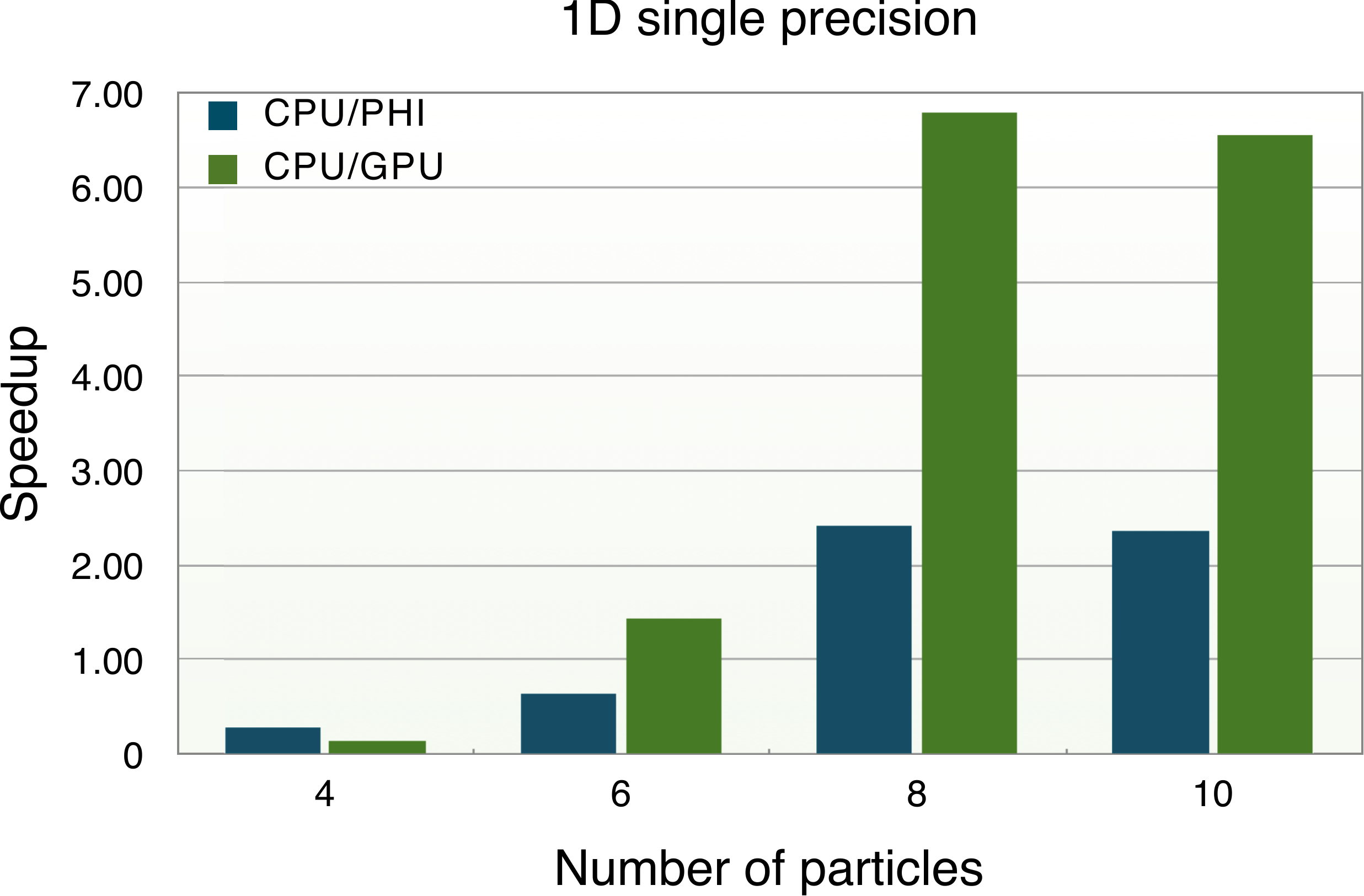}
 \end{subfigure}
\begin{subfigure}{0.5\textwidth}
 \includegraphics[width=0.9\textwidth]{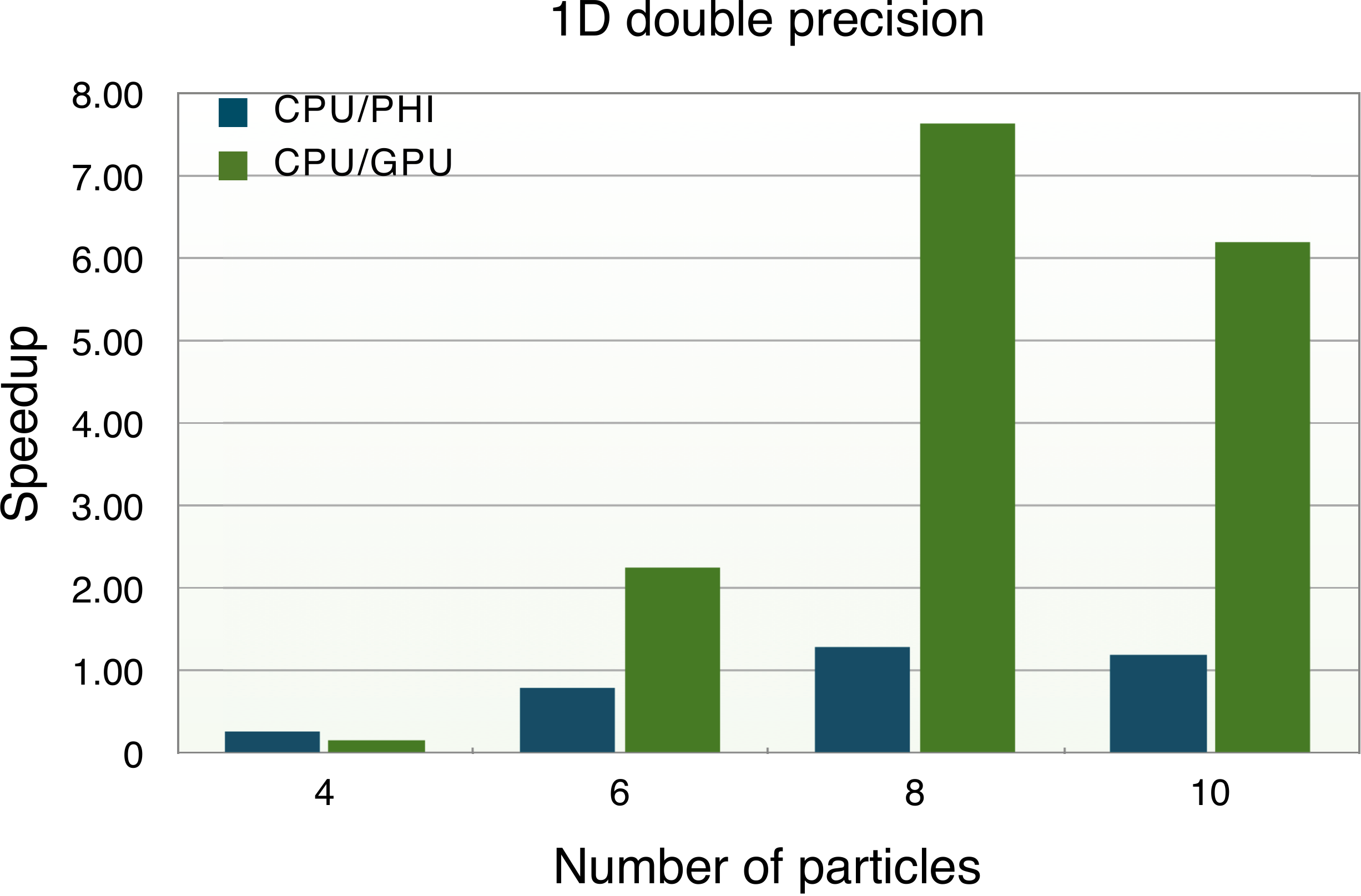}
\end{subfigure}
 \caption{ (Color online) \label{fig1D2} (top row) The execution times of the CPU, the Xeon Phi and the GPU in the 1D lattice with different particle numbers. There are equal numbers of spin up and spin down particles. (bottom row) The speedup factors of the Xeon Phi and the GPU compared to the CPU, computed from the execution times in the top row figures.}  
 \end{figure*}

\begin{figure*}[t.]
\begin{minipage}{\textwidth}
  \begin{minipage}[b]{0.5\textwidth}
    \centering
   \begin{tabular}{|c|c|c|c|}
    \multicolumn{4}{c}{Checkerboard single-precision execution times}\\[0.25cm]
\hline 
\# of particles & CPU(ms) & PHI(ms) & GPU(ms)\tabularnewline
\hline
\hline 
4 & 0.50 & 1.31 & 2.20 \tabularnewline
\hline 
6 & 12.5 & 9.97 & 3.84 \tabularnewline
\hline 
8 & 207 & 87.0 & 40.8 \tabularnewline
\hline 
10 & 2010 & 1090 & 425 \tabularnewline
\hline 
\end{tabular}
  \end{minipage}
  \hfill
  \begin{minipage}[b]{0.5\textwidth}
    \centering
    \begin{tabular}{|c|c|c|c|}
     \multicolumn{4}{c}{Checkerboard double-precision execution times}\\[0.25cm]
\hline 
\# of particles & CPU(ms) & PHI(ms) & GPU(ms)\tabularnewline
\hline
\hline 
4 & 0.58 & 1.60 & 2.28 \tabularnewline
\hline 
6 & 18.6 & 15.3 & 5.17 \tabularnewline
\hline 
8 & 342 & 199 & 74.8 \tabularnewline
\hline 
10 & 3420 & 2790 & 712 \tabularnewline
\hline 
\end{tabular}

        \end{minipage}
  \end{minipage}

% \hfill
 
  \vspace{0.6cm}
\begin{subfigure}{0.5\textwidth}
\centering
 \includegraphics[width=0.9\textwidth]{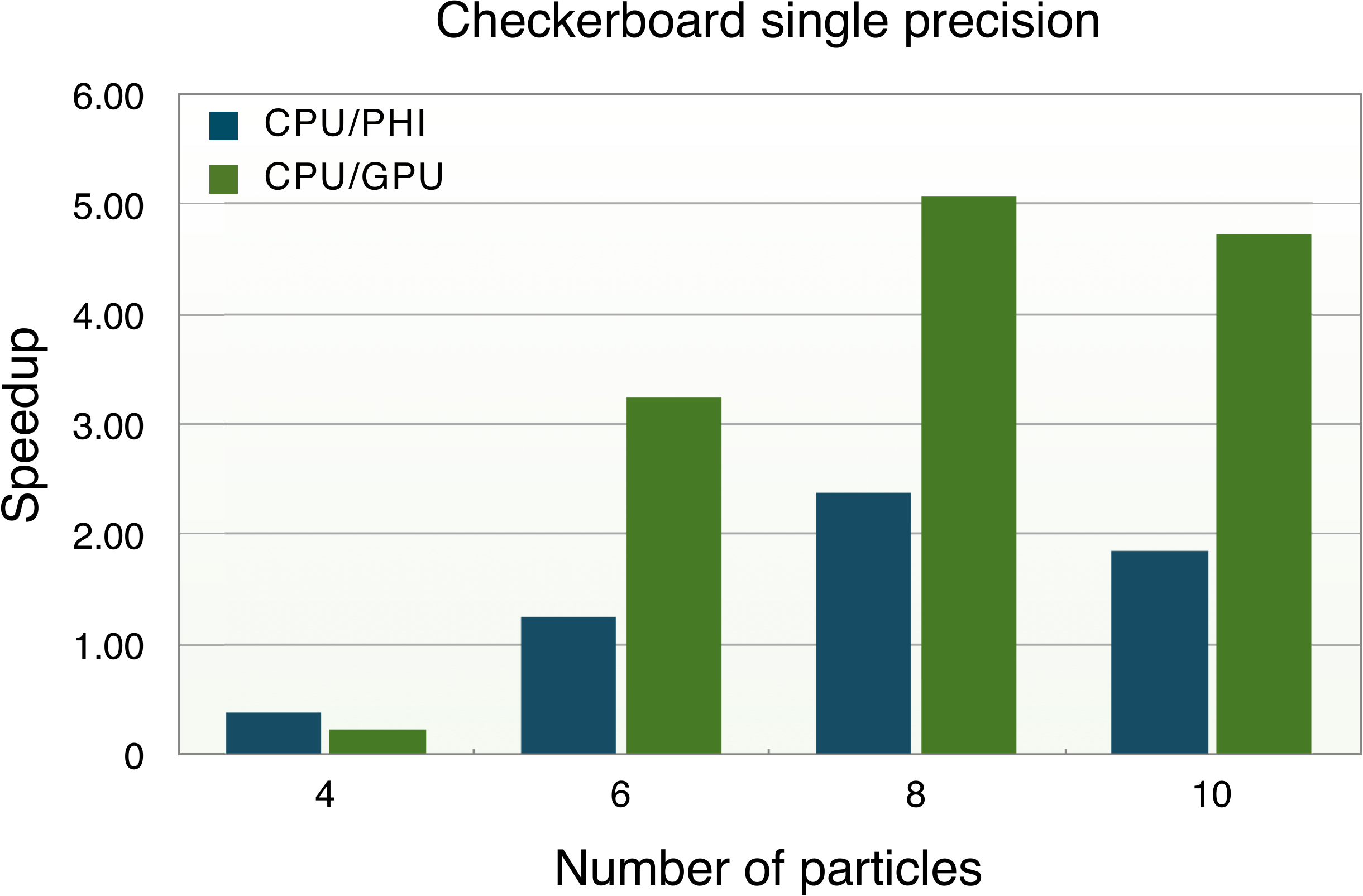}
 \end{subfigure}%
 \begin{subfigure}{0.5\textwidth}
 \includegraphics[width=0.9\textwidth]{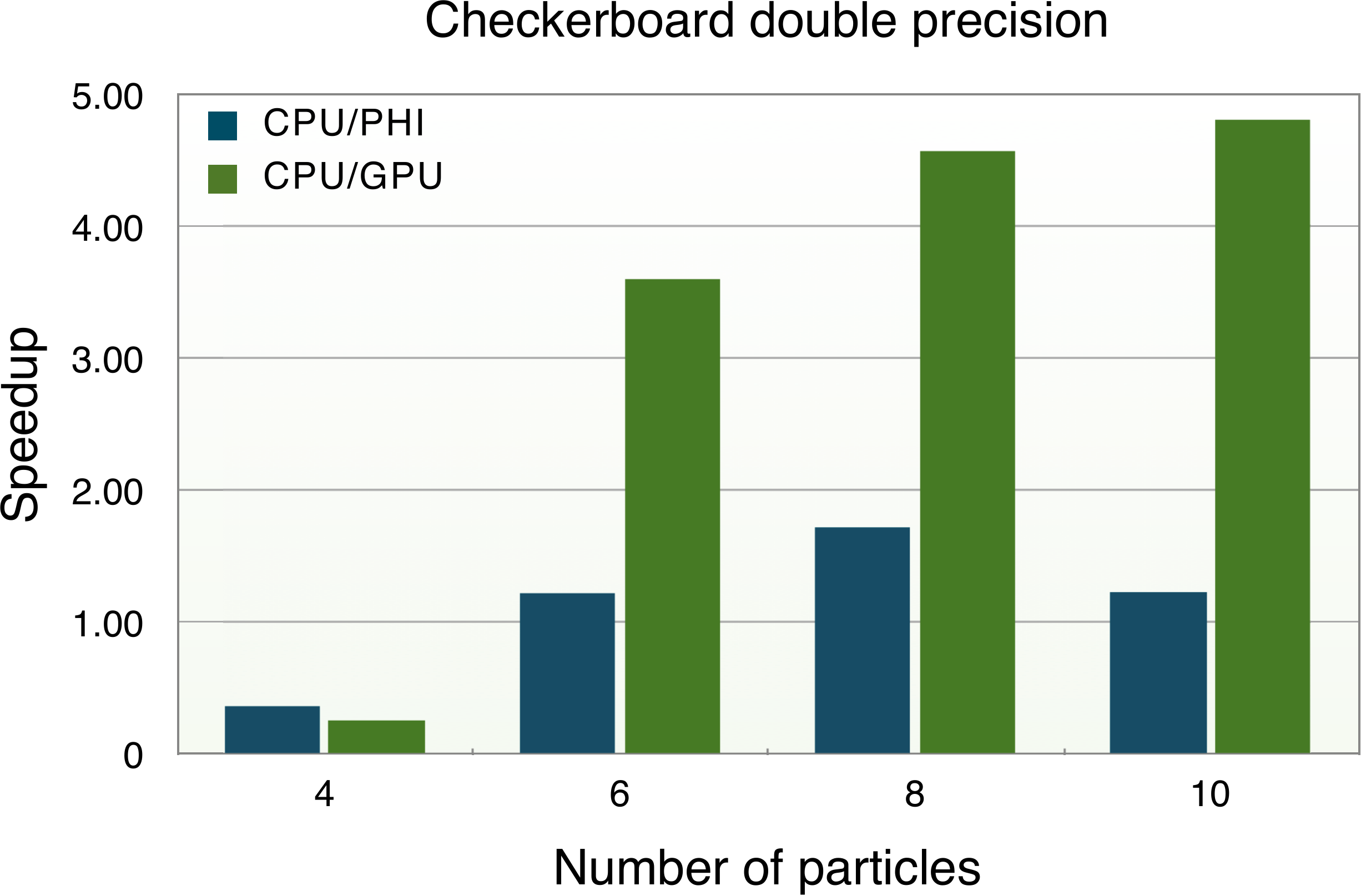}
  \end{subfigure}
  \caption{ (Color online) \label{figCh2} (top row) The execution times of the CPU, the Xeon Phi and the GPU in the checkerboard lattice with different particle numbers. There are equal numbers of spin up and spin down particles. (bottom row) The speedup factors of the Xeon Phi and the GPU compared to the CPU, computed from the execution times in the top row figures.}  
 \end{figure*}

\section{Results}

We benchmarked the performance of the three platforms (CPU, GPU and Xeon Phi) in running the Lancozs algorithm by measuring the execution time of a single iteration of the loop in Algorithm \ref{alg1}. In all tests, we use 12 threads in the CPU and a block size of 256 in the GPU. The thread count in the Xeon Phi is 244 in most cases except the smallest systems with one spin species, where smaller thread counts were found to improve performance. The Xeon Phi was run in native mode in all cases. The results were validated by checking that all three implementations gave the same groundstate energy after 100 Lanczos steps when starting from the same vector. To exclude any initialization overheads and random variation, the execution times were averaged over the 100 steps, excluding the first one. No data transfer has been included in any of the reported times for the Xeon Phi and the GPU.

We present results for two different types of Hamiltonians, each with the 1D and the checkerboard lattices. First, the one spin species Hamiltonians, where all particles have the same spin. This means that we construct the full hopping Hamiltonian matrix and use library implementations for sparse matrix-vector product to operate with the Hamiltonian in the Lanczos algorithm. For the CPU and the Xeon Phi, we use the gemv routine in the MKL library and for the GPU we use the CUSPARSE library. Second, we present results for Hamiltonians with two spin species, where there are an equal number of up and down-spin particles. We form separate hopping matrices for up and down-spin electrons as per Equation \ref{Hsplit} and use the kernels described in Section \ref{pro} to operate with the Hamiltonian. The hopping matrices are stored in the ELL format in all platforms but in the CPU and the Xeon Phi we store them in row-major order and in the GPU in column-major order to enable efficient memory access by the threads. In both one and two spin species cases, the simple axpy, normalization, scaling and dot product operations in the Lanczos algorithm are computed with libraries.

The one spin species results for the 1D and checkerboard lattices are shown in Figures \ref{fig1D1} and \ref{figCh1}, respectively. Looking at the execution times (the top rows), the qualitative behavior is the same in both lattices and with both single and double precision arithmetic. Namely, with a small number of particles, the CPU performs much faster than the accelerators. This is expected because of the small Hilbert space in these cases, leading to insufficient data-parallelism to take advantage of the resources of the Xeon Phi and the GPU. With a larger number of particles, the GPU emerges as clearly the fastest platform with speedups of 2.9 (single) and 4.5 (double) in the 1D lattice and 1.8 (single) and 2.8 (double) in the checkerboard lattice over the CPU. It is, however, noteworthy that in all cases with a large enough particle number the Xeon Phi is faster than than the CPU, reaching speedups of up to 1.4.

Next, we present the results for two spin species in Figures \ref{fig1D2} and \ref{figCh2}.  We present the execution times in a table because of their exponential growth with increasing particle number. The general trend in the two-spin results is the same as with just one spin. The CPU is clearly fastest in the smallest system with four particles, but both accelerators overtake it when the Hilbert space grows larger. Again, the GPU is the clear winner, reaching speedups of up to 7.6 and 5.0 in the 1D and the checkerboard lattice, respectively. The best performance of the Xeon Phi in comparison with the CPU is a speedup of 2.5 in the 1D lattice case with single precision.

To gain insight into the efficiency of our implementation and utilization of the coprocessors, we can compute rough estimates for the achieved floating point performance and memory bandwidth utilization. With one spin species, there is one complex multiplication (6 FLOP) and one complex addition (2 FLOP) per nonzero element of the Hamiltonian in the $Hx$ operation. In addition, the axpy, normalization, scaling and dot product operations in the Lanczos algorithm have a total of $20 \times \dim H$ FLOP. Thus, the number of floating point operations for the single spin species case in one iteration of the Lanczos loop is
$(8 \times \textnormal{numcols} + 20) \times \dim H$,
where numcols is the number of columns in the ELL representation of the Hamiltonian.

In the two spin species case, each nonzero element of $H_{\uparrow}$ is used $\dim H_{\downarrow}$ times and vice versa, so the total number of floating point operations is $(8 \times \textnormal{numcolsUp} + 8 \times \textnormal{numcolsDn} + 20) \times \dim H_{\uparrow} \times \dim H_{\downarrow}$.  

Accurately estimating the sustained memory bandwidth is much harder, since the input vector needs to be transferred multiple times. This is due to the scattered access pattern on the memory in the $Hx$ kernels  and very limited cache sizes for the large systems. For example, in the 18 site checkerboard lattice with 5 up and 5 down spin electrons, one state vector takes around 1.2 GB of memory in double precision, which is huge compared to the 512 kB per core L2 cache on the Xeon Phi and the 1.6 MB shared L2 cache on the GPU. However, we can compute a lower bound by assuming that all the matrices and vectors are transferred only once. In the single spin species case, this gives the amount of transferred memory in an iteration of the Lanczos loop as $((p+4) \times \textnormal{numcols} + 12 \times p) \times \dim H$ bytes, where $p$ is equal to 8 and 16 for single and double precision, respectively. In the two spin species case, it is $(p+4) \times (\textnormal{numcolsUp} \times \dim H_{\uparrow} + \textnormal{numcolsDn} \times \dim H_{\downarrow}) + 12 \times p \times \dim H_{\uparrow} \times \dim H_{\downarrow}$ bytes. 

To find out whether our program is compute or memory bound, we can compute the FLOP/byte ratio from the expressions above. For the one spin species cases the ratio is always below 1. In the two spin species cases, there is a lot more reuse of the data, but the FLOP/byte ratio is still below 5 in all cases except the checkerboard lattice in single precision, where it reaches 10.1 with 10 particles. According to References \cite{Saule2014} and \cite{MICbook}, the practical maximum bandwidth of the Xeon Phi is around 180 GB/s, so with the 2.4 TFLOPS and 1.2 TFLOPS floating point performances in single and double precision, we expect the FLOP/byte balance points to be at around 13 and 7, respectively. For the GPU, the balance points are around 17 for single and 6 for double precision. Even with our lower bound estimate of the memory bandwidth, our application is clearly memory bound.

To achieve the maximum floating point performance on the two coprocessors, the system needs to be very large,  $\dim H \sim 10^6$. Common to both the Xeon Phi and the GPU, when the system size is large, double precision performance is significantly worse than single precision. This indicates that we are mostly limited by the memory bandwidth instead of latency. 

In the four cases with one spin species where we are using the MKL and CUSPARSE libraries for the $Hx$ operation, the maximum performances vary between 10 and 30 GFLOPS for the Xeon Phi, and between 40 and 60 GFLOPS for the GPU. With two spin species, where our custom kernels are used, the maximum performances vary between 15 and 85 GFLOPS for the Xeon Phi, and between 75 and 180 GFLOPS for the GPU. Both coprocessors perform better with the checkerboard lattice. This is probably due to the greater nonzero density compared to the 1D lattice, leading to improved cache usage. As expected, the much larger FLOP/byte ratio in the two spin species case leads to significantly increased performance.

\section{Conclusions}

We have implemented the Lanczos algorithm to compute the ground state energy of a many-particle quantum lattice model on three platforms: a multi-core Intel Xeon CPU, an Intel Xeon Phi coprocessor and an NVIDIA GPU. The CPU and the Xeon Phi were parallelized with OpenMP, and with only one spin species in the model, the MKL library was used to compute the sparse matrix-vector product in the Lanczos algorithm. With two spin species, a custom OpenMP function was used. The GPU was programmed with CUDA. In the single spin species case, we used the CUSPARSE library and with two spin species we used a custom CUDA kernel.

We benchmarked the programs with single and double precision arithmetic in two different lattice geometries: a 1D ring with nearest-neighbour hopping and a checkerboard lattice with hoppings up to the third nearest-neighbor lattice sites. In all cases, the CPU is the fastest of the three platforms when the particle number is very low. With larger particle numbers, the GPU is the fastest, with speedup factors of up to 7.6 compared to the CPU. While the Xeon Phi is never the fastest of the three test platforms, it does outperform the CPU when the particle number is sufficiently high, by up to a speedup of 2.5. This is important, since an existing CPU code can be run on the Xeon Phi with practically no coding effort, resulting in an instant performance gain. All in all, our results indicate that with the current hardware, graphics processors with custom low level kernels offer the best performance in exactly diagonalizing many-particle quantum lattice models at large system sizes. The Xeon Phi was shown to be a good choice for gaining a significant speedup over an existing multi-core code with very little programming effort. 
 
\section*{Acknowledgements}
T.S acknowledges financial support from the Finnish
Doctoral Programme in Computational Sciences FICS. This research has
also been supported by the Academy of Finland through its Centres of
Excellence Program (project no. 251748).  We acknowledge the
computational resources provided by Aalto Science-IT project and
Finland's IT Center for Science (CSC).

%% The Appendices part is started with the command \appendix;
%% appendix sections are then done as normal sections
%% \appendix

%% \section{}
%% \label{}

%% References
%%
%% Following citation commands can be used in the body text:
%% Usage of \cite is as follows:
%%   \cite{key}          ==>>  [#]
%%   \cite[chap. 2]{key} ==>>  [#, chap. 2]
%%   \citet{key}         ==>>  Author [#]

%% References with bibTeX database:

\bibliographystyle{model1-num-names}
\bibliography{Siro}

\begin{thebibliography}{20}
\expandafter\ifx\csname natexlab\endcsname\relax\def\natexlab#1{#1}\fi
\providecommand{\bibinfo}[2]{#2}
\ifx\xfnm\relax \def\xfnm[#1]{\unskip,\space#1}\fi
%Type = Article
\bibitem[{Baity-Jesi et~al.(2014)Baity-Jesi, Fern\'andez, Mart\'{i}n-Mayor, and
  Sanz}]{Intro1}
\bibinfo{author}{M.~Baity-Jesi}, \bibinfo{author}{L.~A. Fern\'andez},
  \bibinfo{author}{V.~Mart\'{i}n-Mayor}, \bibinfo{author}{J.~M. Sanz},
\newblock \bibinfo{title}{Phase transition in three-dimensional {H}eisenberg
  spin glasses with strong random anisotropies through a multi-{GPU}
  parallelization},
\newblock \bibinfo{journal}{Phys. Rev. B} \bibinfo{volume}{89}
  (\bibinfo{year}{2014}) \bibinfo{pages}{014202}.
%Type = Article
\bibitem[{Ambrose and Stamps(2013)}]{Intro2}
\bibinfo{author}{M.~C. Ambrose}, \bibinfo{author}{R.~L. Stamps},
\newblock \bibinfo{title}{{M}onte {C}arlo simulation of the effects of
  higher-order anisotropy on the spin reorientation transition in the
  two-dimensional {H}eisenberg model with long-range interactions},
\newblock \bibinfo{journal}{Phys. Rev. B} \bibinfo{volume}{87}
  (\bibinfo{year}{2013}) \bibinfo{pages}{184417}.
%Type = Article
\bibitem[{Manssen and Hartmann(2015)}]{Intro3}
\bibinfo{author}{M.~Manssen}, \bibinfo{author}{A.~K. Hartmann},
\newblock \bibinfo{title}{Aging at the spin-glass/ferromagnet transition:
  {M}onte {C}arlo simulations using graphics processing units},
\newblock \bibinfo{journal}{Phys. Rev. B} \bibinfo{volume}{91}
  (\bibinfo{year}{2015}) \bibinfo{pages}{174433}.
%Type = Inproceedings
\bibitem[{Harju et~al.(2013)Harju, Siro, Canova, Hakala, and
  Rantalaiho}]{GPUreview}
\bibinfo{author}{A.~Harju}, \bibinfo{author}{T.~Siro}, \bibinfo{author}{F.~F.
  Canova}, \bibinfo{author}{S.~Hakala}, \bibinfo{author}{T.~Rantalaiho},
\newblock \bibinfo{title}{Computational physics on graphics processing units},
\newblock in: \bibinfo{booktitle}{Proceedings of the 11th International
  Conference on Applied Parallel and Scientific Computing}, PARA'12,
  \bibinfo{publisher}{Springer-Verlag}, \bibinfo{address}{Berlin, Heidelberg},
  \bibinfo{year}{2013}, pp. \bibinfo{pages}{3--26}.
%Type = Inproceedings
\bibitem[{Heybrock et~al.(2014)Heybrock, Jo\'{o}, Kalamkar, Smelyanskiy,
  Vaidyanathan, Wettig, and Dubey}]{IntroMIC1}
\bibinfo{author}{S.~Heybrock}, \bibinfo{author}{B.~Jo\'{o}},
  \bibinfo{author}{D.~D. Kalamkar}, \bibinfo{author}{M.~Smelyanskiy},
  \bibinfo{author}{K.~Vaidyanathan}, \bibinfo{author}{T.~Wettig},
  \bibinfo{author}{P.~Dubey},
\newblock \bibinfo{title}{Lattice {QCD} with domain decomposition on {I}ntel
  {X}eon {P}hi co-processors},
\newblock in: \bibinfo{booktitle}{Proceedings of the International Conference
  for High Performance Computing, Networking, Storage and Analysis}, SC '14,
  \bibinfo{publisher}{IEEE Press}, \bibinfo{address}{Piscataway, NJ, USA},
  \bibinfo{year}{2014}, pp. \bibinfo{pages}{69--80}.
%Type = Article
\bibitem[{Lyakh(2015)}]{IntroMIC2}
\bibinfo{author}{D.~I. Lyakh},
\newblock \bibinfo{title}{An efficient tensor transpose algorithm for multicore
  {CPU}, {I}ntel {X}eon {P}hi, and {NVidia} tesla {GPU}},
\newblock \bibinfo{journal}{Comp. Phys. Comm.} \bibinfo{volume}{189}
  (\bibinfo{year}{2015}) \bibinfo{pages}{84 -- 91}.
%Type = Article
\bibitem[{Bernaschi et~al.(2014)Bernaschi, Bisson, and Salvadore}]{IntroMIC3}
\bibinfo{author}{M.~Bernaschi}, \bibinfo{author}{M.~Bisson},
  \bibinfo{author}{F.~Salvadore},
\newblock \bibinfo{title}{Multi-{K}epler {GPU} vs. multi-{I}ntel {MIC} for spin
  systems simulations},
\newblock \bibinfo{journal}{Comp. Phys. Comm.} \bibinfo{volume}{185}
  (\bibinfo{year}{2014}) \bibinfo{pages}{2495 -- 2503}.
%Type = Article
\bibitem[{Siro and Harju(2012)}]{Siro_2012}
\bibinfo{author}{T.~Siro}, \bibinfo{author}{A.~Harju},
\newblock \bibinfo{title}{Exact diagonalization of the {H}ubbard model on
  graphics processing units},
\newblock \bibinfo{journal}{Comp. Phys. Comm.} \bibinfo{volume}{183}
  (\bibinfo{year}{2012}) \bibinfo{pages}{1884 -- 1889}.
%Type = Article
\bibitem[{Hubbard(1963)}]{Hubbard}
\bibinfo{author}{J.~Hubbard},
\newblock \bibinfo{title}{{Electron correlations in narrow energy bands}},
\newblock \bibinfo{journal}{Proc. R. Soc. Lond. A} \bibinfo{volume}{276}
  (\bibinfo{year}{1963}) \bibinfo{pages}{238--257}.
%Type = Article
\bibitem[{Sheng et~al.(2011)Sheng, Gu, Sun, and Sheng}]{Sheng_2011}
\bibinfo{author}{D.~N. Sheng}, \bibinfo{author}{Z.-C. Gu},
  \bibinfo{author}{K.~Sun}, \bibinfo{author}{L.~Sheng},
\newblock \bibinfo{title}{{Fractional quantum {H}all effect in the absence of
  {L}andau levels}},
\newblock \bibinfo{journal}{{Nat. Commun.}} \bibinfo{volume}{{2}}
  (\bibinfo{year}{{2011}}).
%Type = Article
\bibitem[{Yang et~al.(2012)Yang, Sun, and Das~Sarma}]{Yang_2012}
\bibinfo{author}{S.~Yang}, \bibinfo{author}{K.~Sun},
  \bibinfo{author}{S.~Das~Sarma},
\newblock \bibinfo{title}{Quantum phases of disordered flatband lattice
  fractional quantum {H}all systems},
\newblock \bibinfo{journal}{Phys. Rev. B} \bibinfo{volume}{85}
  (\bibinfo{year}{2012}) \bibinfo{pages}{205124}.
%Type = Article
\bibitem[{Siro et~al.(2014)Siro, Ervasti, and Harju}]{Siro_2014}
\bibinfo{author}{T.~Siro}, \bibinfo{author}{M.~Ervasti},
  \bibinfo{author}{A.~Harju},
\newblock \bibinfo{title}{Impurities and {L}andau level mixing in a fractional
  quantum {H}all state in a flat-band lattice model},
\newblock \bibinfo{journal}{Phys. Rev. B} \bibinfo{volume}{90}
  (\bibinfo{year}{2014}) \bibinfo{pages}{165101}.
%Type = Article
\bibitem[{Sun et~al.(2011)Sun, Gu, Katsura, and Das~Sarma}]{Sun_2011}
\bibinfo{author}{K.~Sun}, \bibinfo{author}{Z.~Gu},
  \bibinfo{author}{H.~Katsura}, \bibinfo{author}{S.~Das~Sarma},
\newblock \bibinfo{title}{Nearly flatbands with nontrivial topology},
\newblock \bibinfo{journal}{Phys. Rev. Lett.} \bibinfo{volume}{106}
  (\bibinfo{year}{2011}) \bibinfo{pages}{236803}.
%Type = Article
\bibitem[{Sharma and Ahsan(2015)}]{Sharma}
\bibinfo{author}{M.~Sharma}, \bibinfo{author}{M.~Ahsan},
\newblock \bibinfo{title}{Organization of the {H}ilbert space for exact
  diagonalization of {H}ubbard model},
\newblock \bibinfo{journal}{Comp. Phys. Comm.} \bibinfo{volume}{193}
  (\bibinfo{year}{2015}) \bibinfo{pages}{19 -- 29}.
%Type = Book
\bibitem[{Saad(2003)}]{itermethods}
\bibinfo{author}{Y.~Saad}, \bibinfo{title}{Iterative Methods for Sparse Linear
  Systems}, \bibinfo{publisher}{Society for Industrial and Applied
  Mathematics}, \bibinfo{year}{2003}.
%Type = Manual
\bibitem[{NVIDIA(2015)}]{CUDAguide}
\bibinfo{author}{NVIDIA}, \bibinfo{title}{CUDA C Programming Guide},
  \bibinfo{year}{2015}.
  \bibinfo{note}{{h}ttp://docs.nvidia.com/cuda/cuda-c-programming-guide/}.
%Type = Book
\bibitem[{Jeffers and Reinders(2013)}]{MICbook}
\bibinfo{author}{J.~Jeffers}, \bibinfo{author}{J.~Reinders},
  \bibinfo{title}{Intel {X}eon {P}hi Coprocessor High Performance Programming},
  \bibinfo{publisher}{Morgan Kaufmann Publishers Inc.}, \bibinfo{address}{San
  Francisco, CA, USA}, \bibinfo{edition}{1st} edition, \bibinfo{year}{2013}.
%Type = Manual
\bibitem[{NVIDIA(2016)}]{Pascal}
\bibinfo{author}{NVIDIA}, \bibinfo{title}{Pascal Architecture Whitepaper},
  \bibinfo{year}{2016}.
  \bibinfo{note}{{h}ttp://images.nvidia.com/content/pdf/tesla/whitepaper/pascal-architecture-whitepaper.pdf}.
%Type = Inproceedings
\bibitem[{Liu et~al.(2013)Liu, Smelyanskiy, Chow, and Dubey}]{MICspmv}
\bibinfo{author}{X.~Liu}, \bibinfo{author}{M.~Smelyanskiy},
  \bibinfo{author}{E.~Chow}, \bibinfo{author}{P.~Dubey},
\newblock \bibinfo{title}{Efficient sparse matrix-vector multiplication on
  x86-based many-core processors},
\newblock in: \bibinfo{booktitle}{Proceedings of the 27th International ACM
  Conference on International Conference on Supercomputing}, ICS '13,
  \bibinfo{publisher}{ACM}, \bibinfo{address}{New York, NY, USA},
  \bibinfo{year}{2013}, pp. \bibinfo{pages}{273--282}.
%Type = Inbook
\bibitem[{Saule et~al.(2014)Saule, Kaya, and {\c{C}}ataly{\"u}rek}]{Saule2014}
\bibinfo{author}{E.~Saule}, \bibinfo{author}{K.~Kaya},
  \bibinfo{author}{{\"U}.~V. {\c{C}}ataly{\"u}rek}, \bibinfo{title}{Parallel
  Processing and Applied Mathematics: 10th International Conference, PPAM 2013,
  Warsaw, Poland, September 8-11, 2013, Revised Selected Papers, Part I},
  \bibinfo{publisher}{Springer Berlin Heidelberg}, \bibinfo{address}{Berlin,
  Heidelberg}, pp. \bibinfo{pages}{559--570}.

\end{thebibliography}

%% Authors are advised to submit their bibtex database files. They are
%% requested to list a bibtex style file in the manuscript if they do
%% not want to use model1-num-names.bst.

%% References without bibTeX database:

% \begin{thebibliography}{00}

%% \bibitem must have the following form:
%%   \bibitem{key}...
%%

% \bibitem{}

% \end{thebibliography}

\end{document}